\documentclass[a4paper, 12pt, parskip=full-]{scrartcl}
\usepackage[top = 2.5cm, bottom = 2.5cm, left = 2cm, right = 2cm, footskip = 0.6cm]{geometry}

\usepackage[english]{babel}								% (für englische Silbentrennung)
\usepackage[T1]{fontenc}										% (Verwendung von erweiterten Zeichensätzen der Schriftarten und für bessere Silbentrennung)
\usepackage[lighttt]{lmodern}							% (damit bei Verwendung des Packets "fontenc" die Schrift nicht pixelig wird + Option die für das Einfügen von Source Code mit dem listings-Paket nötig ist)

\usepackage[pdftex]{graphicx}				% (zum Einbinden von Grafiken)
\usepackage[skip=0pt]{subcaption}							% (z.B. wenn mehrere Bilder, Tabellen, usw. in einer Umgebung eigene Beschriftungen bekommen sollen)

\usepackage{amsmath}						% (für verbesserte Darstellung von mathematischen Formeln)
\usepackage{mathrsfs}							% (z.B. für die Schriftart "\mathscr")
\usepackage{fixmath}								% (z.B. für kursive griechische Großbuchstaben)
\usepackage{upgreek}							% (für NICHT kursive griechische Kleinbuchstaben, z.B. \upmu)
\usepackage{units}										% (für den richtigen Abstand zwischen Wert und Einheit (\unit[1]{m}) und Einheitenbrüche (\unitfrac[1]{m}{s}) im Text)   -   funktioniert nicht in [] von \caption

\usepackage{remreset}							% (um automatische Zähler-Resets in LaTeX zu deaktivieren)
\makeatletter
\@removefromreset{footnote}{chapter}
\makeatother

\usepackage[bottom, flushmargin, hang]{footmisc}

\usepackage[backend=bibtex, hyperref=true, backref=false, sorting=none, style=numeric-comp, minbibnames=2, mincitenames=2, maxbibnames=5, maxcitenames=5, maxitems=5]{biblatex}
\bibliography{bibliography}							% (Einbinden der BibTeX-Datei bibliography.bib)
\usepackage[babel, english=british]{csquotes}				% (wird für BibTeX benötigt)
\usepackage{xcolor}											% (für Textfarben, z.B. \color{red} oder \textcolor{red}{Text})
\usepackage{float} 												% (um das floating von Bildern oder Tabellen unterdrücken zu können (mit \begin{figure/table}[H]))
\usepackage{threeparttable} 					% (für Tabellen "Fußnoten" - funktioniert nicht zusammen mit "tabu")
\definecolor{darkblue}{rgb}{0,0,1}
\usepackage[breaklinks=true, colorlinks=true, linkcolor=darkblue, urlcolor=darkblue, citecolor=darkblue]{hyperref}			% (für Verweise innerhalb des Dokuments oder ins Internet)
\begin{document}

	%
	% einfügen der einzelnen Abschnitte
	%

	% Titelseite
	\pagenumbering{alph}           % damit jede Seite eine eindeutige Seitenzahl besitzt und "hyperref" funktioniert
	\newgeometry{left = 3.8cm, right = 3.8cm} % ändert die Seitengeometrie ab hier
	\pagestyle{empty}
	\vspace*{1cm}
\begin{center}
	\huge
	\textbf{Peculiarities in the Simulation \\of Optical Physics with Geant4} \\[1cm]
	\normalsize
	%\textbf{Erik Dietz-Laursonn}, Thomas Hebbeker, Markus Merschmeyer\\[0.5cm]
	Erik Dietz-Laursonn\\[0.5cm]
	\small
	III. Physikalisches Institut A, RWTH Aachen University, \\
	Otto-Blumenthal-Str., 52056 Aachen, Germany
\end{center}

	% Zusammenfassung
	\vspace*{2cm}
\paragraph*{Abstract}

% 1/4 - 1/3 Problembeschreibung (3 wichtigste Probleme)
% 2/4 Was habe ich / hat meine Arbeit zur Lösung veigetragen?
% 1/4 Was ist dabei herausgekommen?

Geant4 is a complex and widely-used software toolkit for the simulation of the passage of particles through matter and the interactions they undergo. 
It contains very extensive and flexible optical physics capabilities.
These allow for the specification of a lot of properties for materials and surfaces as well as for the optical physics process itself.
%Therefore, a variety of properties can be specified for materials and surfaces as well as for the optical physics process itself.

Because of this large variety of possible adjustments,
there is a number of peculiarities of the optical physics in Geant4, which the user has to be aware of in order to avoid incorrect simulation results due to user mistakes.
As these peculiarities are not always easy to recognise, this can be a serious problem, especially for less experienced users of Geant4.

%In order to assist users of optical physics in Geant4 in this intention, this paper gives a summary of peculiarities of the optical physics in Geant4.
In order to assist users of optical physics in Geant4 in avoiding mistakes, this paper gives a summary of peculiarities of the optical physics in Geant4.
	\newpage

	% Kapitel
	\restoregeometry               % stellt wieder die im document-header vorgegebene Seitengeometrie her
	\pagenumbering{arabic}         % damit das Hauptdokument arabisch nummeriert wird
	\pagestyle{plain}

\section{Introduction}

\vspace{-0.15cm}
Geant4~\cite{Geant, Geant4_a_simulation_toolkit, Geant4_developments_and_applications} is a \texttt{C++}\mbox{-}based software toolkit for the simulation of the passage of particles through matter and the interactions they undergo. Geant4 can cover many physics processes for electromagnetic and hadronic physics for a large energy range from the eV to the TeV scale. Additionally, Geant4 contains very extensive and flexible optical physics capabilities.

\vspace{-0.05cm}
In Geant4, the optical physics has an exceptional position among the physics processes, as it adds special particles (optical photons) and new properties for materials and optical surfaces. Being the only particle that can be reflected or refracted at optical surfaces%\footnote{In Geant4, optical photons are particles, i.e.~single photons can only be either reflected or refracted (or absorbed in case of a surface between a dielectric and a metal). To obtain the splitting of a light beam according to the reflectivity/transmittance of the optical surface, several optical photons have to be simulated.}
, as well as being only created in optical processes like scintillation, Cherenkov radiation, and wavelength-shifting (WLS) makes the \texttt{G4OpticalPhoton} differing from the ``usual'' high-energy particle-physics photon (\texttt{G4Gamma}) in Geant4. Optical properties need to be assigned to the materials whenever optical physics processes are to be considered in the simulation. Every material needs at least a refractive index spectrum (which corresponds to the dispersion relation) and an attenuation length spectrum, though the attenuation length is by default set to infinity if it is not defined. Special optical materials, i.e.~scintillating and WLS materials, additionally require the specification of the emission spectra as well as of the rise and decay times. More properties can be assigned to optical surfaces between volumes, e.g.~the reflectivity of the surface.

\vspace{-0.1cm}
\section{Peculiarities of the Optical Physics of Geant4}

\vspace{-0.15cm}
The fact that the optical physics in Geant4 is designed to be very flexible and extensive leads to some peculiarities, which the user has to be aware of in order to avoid incorrect simulation results due to user mistakes. As these peculiarities are not always easy to recognise, this can be a serious problem, especially but not exclusively for less experienced users of Geant4. In order to assist users of optical physics in Geant4 in handling its peculiarities, they will be summarised in the following list (which is not to be considered as complete).

\vspace{-0.1cm}
\subsection{General Peculiarities}

\begin{itemize}
	\item
	\vspace{-0.25cm}
	All optical properties are specified as a function of the photon energy within Geant4 and have to be sorted by rising energy values. Thus, if a property spectrum is available as a function of the photon wavelength and/or sorted by falling energy values, the user has to convert the spectrum into a function of rising photon energy, before passing it to Geant4. In this context, the user has to be aware that the spectra are also processed as a function of the photon energy, e.g.~when interpolating between two points of the spectrum. Additionally, all optical properties have to be specified on the same energy range. Otherwise errors will occur, e.g.~if a scintillation photon is created with an energy for which no refractive index is specified in the material.
	\item
	\vspace{-0.05cm}
	The rise time is a property of the scintillation process, which are decisive for the correct simulation of the signal shape/timing of the detector. The rise time variables of the scintillation process are deactivated by default, i.e.~the simulated rise times are always zero, independent of the user\mbox{-}specified rise time values of the materials. If finite rise times are to be used, the rise time variables have to be activated when registering the optical physics process in the \texttt{PhysicsList}. A code example demonstrating this procedure can be found in~\cite{DietzLaursonn_Diss}.
	\item
	\vspace{-0.05cm}
	The decay time spectrum of the WLS process (also essential for the accurate simulation of the signal shape/timing) is by default a $\delta$\mbox{-}function rather than an exponential spectrum. If an exponential spectrum is to be used, this has to be activated when registering the optical physics process in the \texttt{PhysicsList} (for code example cf.~\cite{DietzLaursonn_Diss}).
	\item
	\vspace{-0.05cm}
	The spectra of the reflectivity, the refractive index, and the attenuation length are interpolated by a straight line between two given points of the spectrum. In contrast, emission spectra are interpolated by applying the mean of the values of two neighbouring points as constant value between these points. This can be shown with a simple simulation, cf.~figure~\ref{fig:validationOfMaterialPropertyInputAndPhysicsProcesses_emissionSpectra}. In order to avoid unexpected behaviour, the latter should therefore be specified with a high density of data points.
\begin{figure}[!b!]	% Positionierung an der Unterkante einer Seite
	\vspace{-0.4cm}	% Fließtext kann näher an die obere Kante der Bilder herankommen, die Bilder selbst verschieben sich nicht
	\centering
	\includegraphics[width=7.92cm]{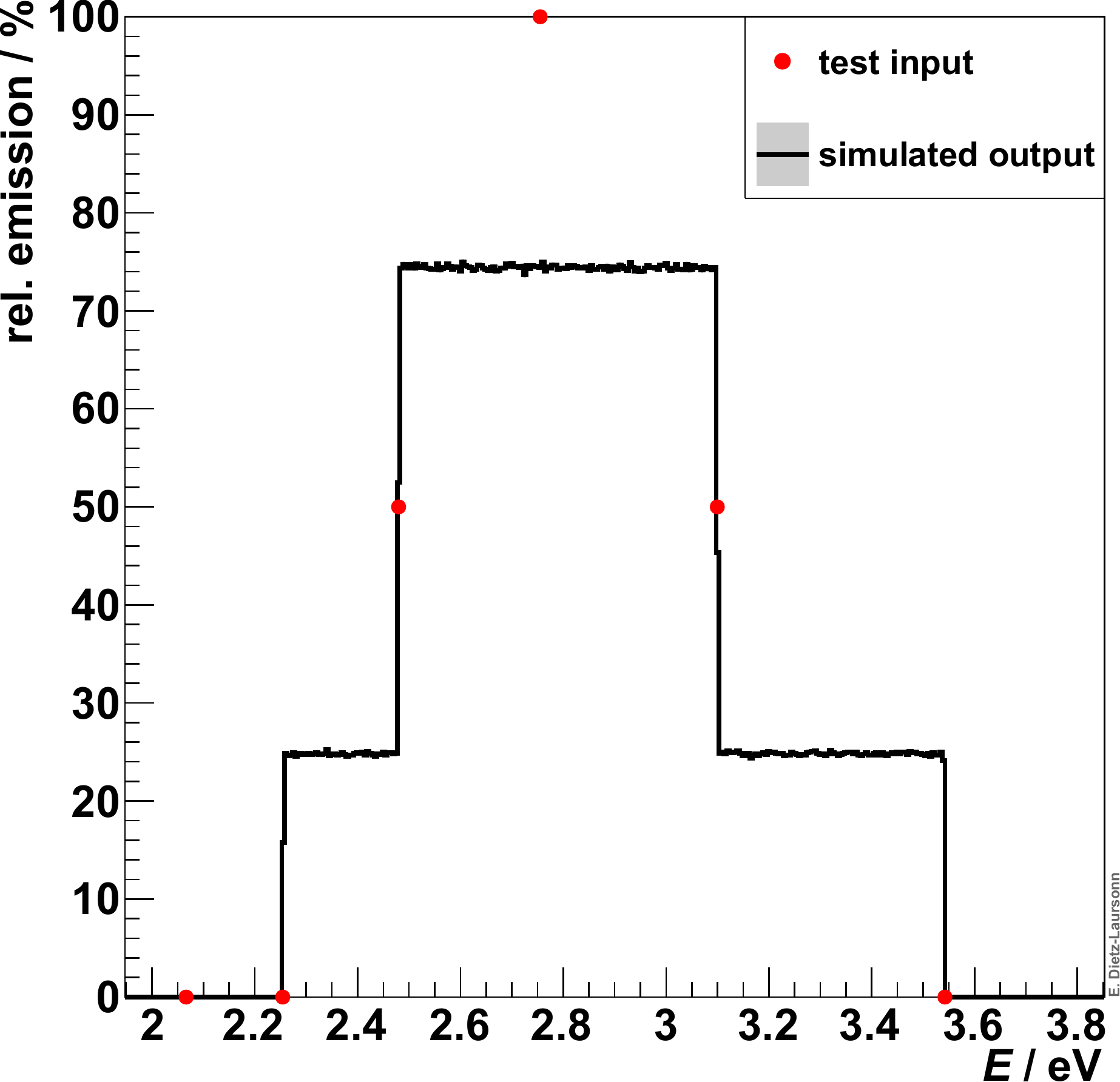}
	
	\vspace{-0.1cm}	% Abstand zwischen Bildern und Caption
  \caption{Test of the processing of (relative) emission spectra in Geant4, showing that simulated emission spectra are not interpolated by a straight line between two points of the spectrum. The dots represent the emission spectrum input used for simulated scintillating material, the lines illustrate the initial energy of the corresponding scintillation photons.}
	\label{fig:validationOfMaterialPropertyInputAndPhysicsProcesses_emissionSpectra}
	\vspace{-0.2cm}	% Bild rückt weiter nach unten (außer vor footnotes)
\end{figure}
	\item
	\vspace{-0.05cm}
	By default, the processing of a (mother) particle is postponed in Geant4 at the point of time when it creates new \texttt{G4OpticalPhoton}s (e.g.~via scintillation)~\cite{Geant4HyperNews_tracks}. After the processing of the \texttt{G4OpticalPhoton}s (together with all their potential child particles) has been completed, the mother particle's processing will continue. Therefore, it may appear to the user, as if the mother particle's trajectory consists of multiple \texttt{G4Track}s. This can lead to double counting or only considering one part of the trajectory of the mother particle, when not considering the track identification number, which is unique to all particles and thus the same for ``all \texttt{G4Track}s'' of the mother particle's trajectory. Whereas also the practice of preponing \texttt{G4OpticalPhoton}s can be changed when registering the optical physics process in the \texttt{PhysicsList} (for code example cf.~\cite{DietzLaursonn_Diss}), this approach is not recommended as it can lead to problems with memory allocation for large numbers of \texttt{G4OpticalPhoton}s~\cite{Geant4HyperNews_tracks}.
	\item
	\texttt{G4SensitiveDetector}s are not only activated by optical photons that enter the corresponding volume, but also by optical photons that are reflected at the surface of this volume. This is simply due to the way how sensitive detectors and optical reflections are treated internally by Geant4 and has to be considered by the user.
\end{itemize}

%%%%%%%%%%%%%%%%%%%%%%%%%%%%%%%%%%%%%%%%%%

\vspace{-0.05cm}
\subsection{Cherenkov Radiation}

\vspace{-0.05cm}
Cherenkov radiation is emitted by charged particles traversing matter, if their velocity $v$ exceeds the effective speed of light in the material ($c / n$, where $n$ is the real part of the material's refractive index). This leads to the Cherenkov criterion 
\vspace{-0.1cm}
\begin{equation}
	\beta = \frac{v}{c} > \frac{1}{n}.
	\vspace{-0.1cm}
\end{equation}
The mean number of Cherenkov photons that is produced by a particle with charge $z$ per path length $x$ and photon energy $E_\gamma$ or photon (vacuum) wavelength $\lambda$ is given by~\cite{PDG12}
\begin{equation}
	\frac{\text{d}^2 N}{\text{d} x \, \text{d} E_\gamma} = \frac{\alpha z^2}{\hbar c} \left( 1 - \frac{1}{\beta^2 n^2(E_\gamma)} \right)
	\hspace{0.5cm} \text{or} \hspace{0.5cm}
	\frac{\text{d}^2 N}{\text{d} x \, \text{d} \lambda} = \frac{2 \pi \alpha z^2}{\lambda^2} \left( 1 - \frac{1}{\beta^2 n^2(\lambda)} \right).%   so auch in Geant
	\label{equ:interactions_electromagneticInteractionsOfMassiveParticlesInMatter_radiativeEnergyLoss_CherenkovPhotons}
\end{equation}
Here, $\alpha$ is the fine\mbox{-}structure constant. Whereas this equation does not restrict Cherenkov radiation with respect to energy or wavelength, it is predominantly observed in the optical range.

For Geant4 simulations taking Cherenkov radiation into account, the following things have to be considered:
\begin{itemize}
	\item
	\vspace{-0.05cm}
	Geant4 itself does not limit the emission spectrum of the Cherenkov process to reasonable energy ranges but creates Cherenkov photons distributed over the full energy range for which $n$ is specified (and the Cherenkov criterion is fulfilled). Thus, simulated Cherenkov radiation can occur in non\mbox{-}physical energy regions, e.g.~in the keV, MeV or TeV range, instead of in the optical range. Therefore, the refractive index should be restricted to the necessary optical energy range or the physical Cherenkov energy range, respectively. As explained before, all other optical properties should be restricted to the same energy range, since defining individual optical properties on different energy ranges will most probably cause problems.
	\item
	\vspace{-0.05cm}
	Even though the refractive index spectra are usually interpolated by a straight line between two given points of the spectrum, they should be specified with a high density of data points, as long as they are not constant. Otherwise, unexpected deviations for the number of created Cherenkov photons may occur, as in this context, $1 / n^2$ is calculated for each point of the spectrum and this value is interpolated by a straight line, in contrast to the usual procedure.
	\item
	\vspace{-0.05cm}
The refractive index value corresponding to the highest energy of the spectrum ($n(E_\text{max})$) always has to be the maximum refractive index of the spectrum.
Otherwise, non-physical results will occur. This is caused by two facts:
First, instead of the full refractive index spectrum, only the value $n(E_\text{max})$ is used within Geant4 to examine whether Cherenkov photons are created at all.
Therefore, not a single Cherenkov photon is created for $\beta\gamma$ of the traversing particle of 1 in case of $n_2$ (cf.~figure~\ref{fig:validationOfMaterialPropertyInputAndPhysicsProcesses_Cherenkov}).
This procedure is used to reduce the computing time.
Second, $n(E_\text{max})$ also seems to serve as an upper limit on the refractive index spectrum when it is used to determine the energy distribution of the Cherenkov photon. Therefore, the simulated spectrum for $n_2$ and $\beta\gamma = 3$ is constant instead of decreasing with decreasing $n$ (cf.~figure~\ref{fig:validationOfMaterialPropertyInputAndPhysicsProcesses_Cherenkov}).
\begin{figure}[!t!]	% Positionierung an der Oberkante einer Seite
	\centering
	\includegraphics[width=7.4cm]{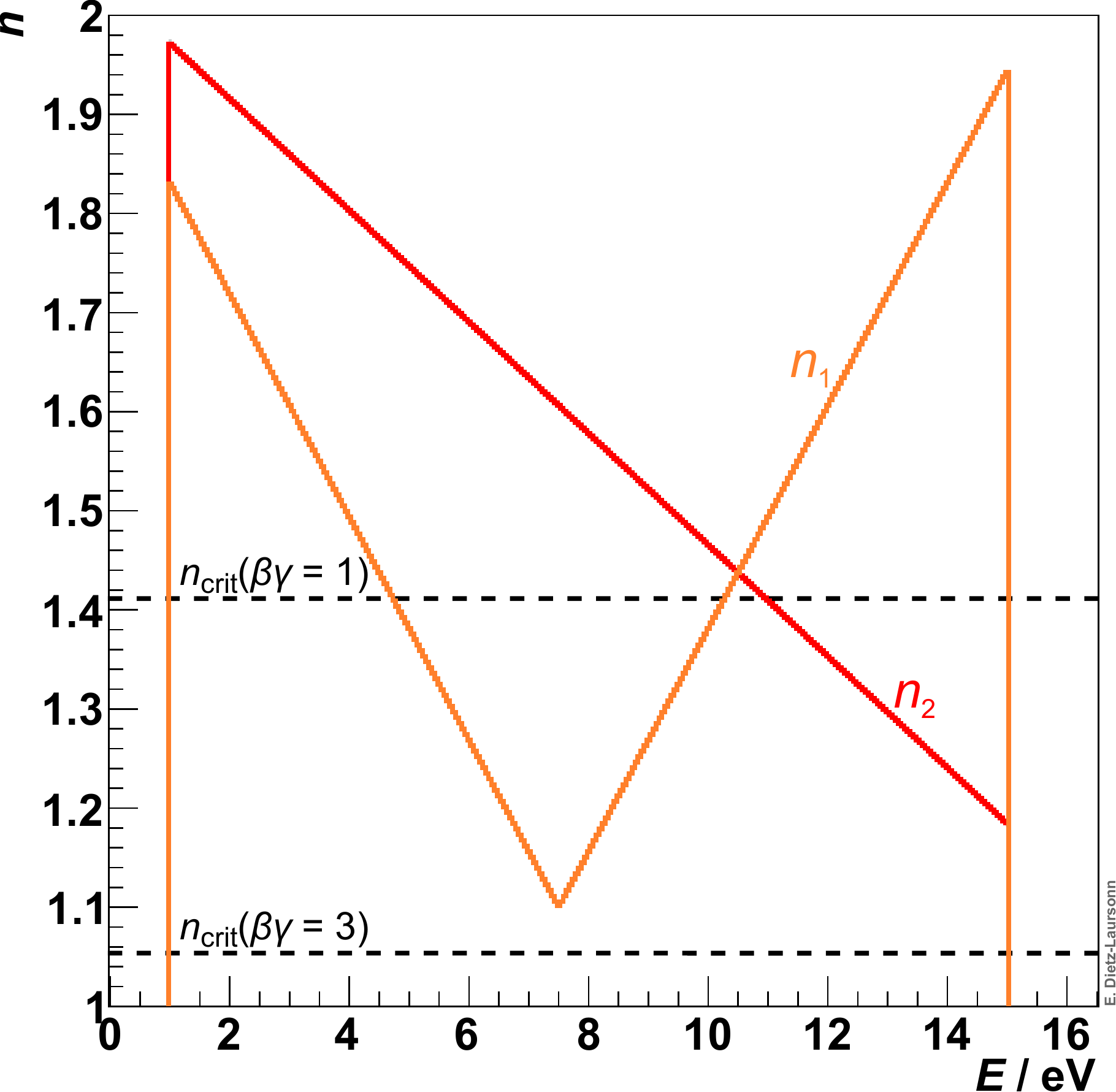}
	\hspace{0.5cm}
	\includegraphics[width=7.4cm]{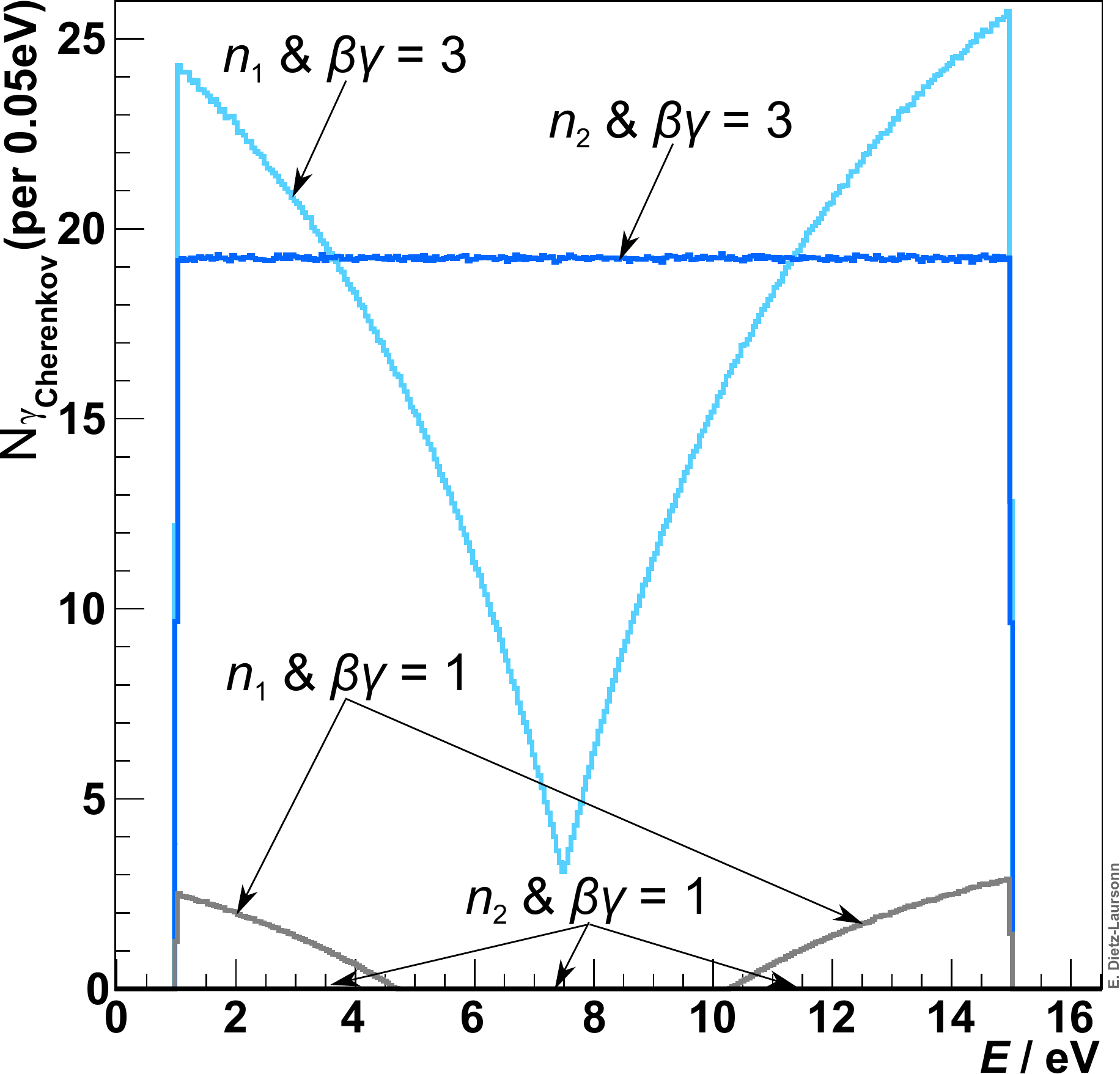}

	\caption{Simulated Cherenkov spectra in case of a linear refractive index with ($n_1$) and without ($n_2$) dip: energy dependence of the refractive index (left) as specified for the simulations and energy spectra of Cherenkov photons (right) as obtained from the simulations. The simulated material layer was \unit[2]{cm} thick.}
  \label{fig:validationOfMaterialPropertyInputAndPhysicsProcesses_Cherenkov}
	%\vspace{-0.3cm}	% Fließtext kann näher an die untere Kante der Bilder herankommen, die Bilder selbst verschieben sich nicht
\end{figure}
	\item
	Another peculiarity occurs for refractive index spectra with a dip.\footnote{An example of a material with dips in the real part of the refractive index is water for the energy range between \unit[0.1]{eV} and \unit[0.5]{eV}~\cite{Refractive_index_of_water}, which is far off the optical range, of course.} At the first glance, the results from a simulation with a linear refractive index spectrum with a dip seem to meet the expectation: In contrast to the case of $\beta\gamma = 3$, the Cherenkov spectrum does not cover the full energy range anymore for the case of $\beta\gamma = 1$ (cf.~figure~\ref{fig:validationOfMaterialPropertyInputAndPhysicsProcesses_Cherenkov}) and the energy ranges without Cherenkov photons match the part of the refractive index spectrum that is below the critical value of $n_\text{crit} = 1 / \beta = \sqrt{2}$. Additionally, the simulated number of Cherenkov photons agrees with the expected value in case of $\beta\gamma = 3$.

But at the same time, for $\beta\gamma = 1$, the number of Cherenkov photons is much lower than expected. The deviation is caused by the fact that the number of Cherenkov photons is calculated with respect to the full energy range in Geant4 instead of considering the energy ranges with $n \geq 1 / \beta$ only. In Geant4, the energy range for the calculation of the total number of Cherenkov photons is only limited, if the refractive index is below the critical value of $1 / \beta$ at the low energy edge of the refractive index spectrum.
\end{itemize}

Summarising the different points, the Cherenkov spectrum meets the expectation as long as the material is normal-dispersive (this is what the Cherenkov radiation in Geant4 was designed for~\cite{Geant4BugReport}) and the non-constant parts of refractive index spectra are specified with a high density of data points. Furthermore, all optical properties should be restricted to the necessary optical energy range or the physical Cherenkov energy range, as this is not done by Geant4 itself.

%%%%%%%%%%%%%%%%%%%%%%%%%%%%%%%%%%%%%%%%%%

\subsection{Birks Reduction of the Scintillation Light Yield}

\vspace{-0.05cm}
According to Birks' widely used semi\mbox{-}empirical model~\cite{Birks}, the actual light yield of organic scintillators is reduced because of recombination and quenching effects between the excited molecules. As these effects have more influence if the density of excited molecules is higher, the response of organic scintillators is not perfectly proportional to the energy deposition of the traversing charged particle. The mean number of photons $N_\gamma$ per path length $x$ of the charged particle's trajectory within the scintillating material can be approximated with Birks' equation~\cite{Birks}
\vspace{-0.05cm}
\begin{equation}
	\frac{\text{d} N_\gamma}{\text{d} x} = \frac{\mathscr{L} \cdot \frac{\text{d} E}{\text{d} x}}{1 + C_\text{B} \cdot \frac{\text{d} E}{\text{d} x}}.
\end{equation}
Here, $\mathscr{L}$ is the light yield at a low density of excited molecules (the value that is typically specified by the manufacturers), 
$\text{d} E / \text{d} x$ is the particle's mean energy loss due to excitation, and $C_\text{B}$ is Birks' constant, which has to be determined experimentally for each scintillating material (cf.~e.g.~\cite{BirksMeasurement}).
For the plastic scintillator BC\mbox{-}408~\cite{BC404_BC408} with $C_\text{B} = \unit[0.0115]{\frac{\text{g}}{\text{MeV} \text{cm}^2}}$~\cite{BirksMeasurement}, typical values of $1 / \left( 1 + C_\text{B} \left( \frac{\text{d} E}{\text{d} x} \right)_\text{ion.} \right)$ are \unit[91.7]{\%}, \unit[97.8]{\%}, and \unit[96.7]{\%} for muons with $E_\text{kin} = \unit[10]{MeV}$%7.917 MeV cm^2 / g
, $E_\text{kin} = \unit[325]{MeV}$ (minimum ionising particle (MIP))%1.956 MeV cm^2 / g
, and $E_\text{kin} = \unit[1.19]{TeV}$ ($E_\text{crit}$)%3 MeV cm^2 / g
, respectively.

For the application of Birks' equation for Geant4 simulations, the following things have to be considered:
\begin{itemize}
	\item
	\vspace{-0.1cm}
	Birks' equation is implemented in Geant4, but Birks' constant is the only optical material property that cannot be specified via a property variable. Instead, it has to be defined directly for the ionisation process of the material. A code example demonstrating this procedure can be found in~\cite{DietzLaursonn_Diss}.
	\item
	\vspace{-0.05cm}
	When considering saturation effects according to Birks' equation in simulations with Geant4, the scintillation light yield qualitatively meets the expectations (reduction). Nevertheless, the quantitatively correct simulation as well as the quantitative comparison of simulation results and measurements is rather complex. The reason for this is the non-linearity of Birks' equation with respect to the energy deposition: higher energy depositions per path length ($\text{d} E / \text{d} x$) undergo larger reductions. As a mathematical consequence, the resulting correction is larger when calculating it for several path segments with fluctuating $\text{d} E / \text{d} x$ than when calculating it for the full path with the mean $\text{d} E / \text{d} x$.
	
	\vspace{-0.05cm}
	Additionally, the energy dependence of Birks' equation leads to another effect: Theoretically, Birks' constant $C_\text{B}$ is a material constant, i.e.~does not depend on the geometry, the way of its determination, etc.. In contrast, the measured values of $C_\text{B}$ do depend on the way they were determined. This becomes clear using the example of the measurements in the original work of Birks~\cite{Birks}. There, the number of scintillation photons of particles that were absorbed within the scintillating material was plotted against $\text{d} E / \text{d} x$-values that were calculated from the initial particle energies via the Bethe-Bloch equation. Obviously, a $C_\text{B}$-value determined from these data would probably not be applicable to particles that are not absorbed, as $\text{d} E / \text{d} x$ is higher shortly before the absorption than at the beginning of the path.

	\vspace{-0.05cm}
	Consequently, the implementation of saturation effects according to Birks' equation into simulations as well as the verification of the correctness of the simulation results is much more complex than it seems to be at first glance. 
	In Geant4, the number of scintillation photons is calculated for every \texttt{G4Step} from its length and from the corresponding energy deposition. 
	This corresponds to values of $C_\text{B}$ that were measured for MIPs (far from being absorbed during the measurement) and have been corrected for the difference between the step length in Geant4 and the effective path length during the measurement. 
	Values of $C_\text{B}$ that were determined differently, will probably lead to incorrect simulation results.
\end{itemize}

%%%%%%%%%%%%%%%%%%%%%%%%%%%%%%%%%%%%%%%%%%

%\vspace{-0.05cm}
\subsection{G4OpticalSurfaces}

%\vspace{-0.05cm}
\texttt{G4OpticalSurface}s can be used to simulate other optical surfaces than perfectly smooth surface between two dielectrics. Using \texttt{G4OpticalSurface}s, several surface models, surface types, and surface finishes can be specified.

%\vspace{-0.1cm}
With respect to the application of \texttt{G4OpticalSurface}s in Geant4 simulations, the following things have to be taken into account:
\begin{itemize}
	\item
	%\vspace{-0.1cm}
	The default reflection model of \texttt{G4OpticalSurface}s differs from the case without \linebreak
	\texttt{G4OpticalSurface}s: If no \texttt{G4OpticalSurface} is defined, the reflection is simulated as geometric reflection at a perfectly smooth optical surface, i.e.~applying Snell's law. When introducing a \texttt{G4OpticalSurface} without specifying the parameters of the reflection model, the default reflection model is diffuse reflection.
	\item
	%\vspace{-0.1cm}
	Optical surfaces can be defined via two ways: Between two specific volumes (\texttt{G4Logical}-\texttt{BorderSurface}) or around one volume (\texttt{G4LogicalSkinSurface}). Using \texttt{G4Logical}-\texttt{BorderSurface}s, the two volumes forming the surface have to be specified as well as its properties and the photon direction of flight for which the properties are to be applied. If a \texttt{G4LogicalSkinSurface} is specified, its properties are applied to every surface of the corresponding volume for both directions of flight of the photons. If the properties are not to be applied for one of the surfaces, additional \texttt{G4LogicalBorderSurface}s have to be specified for the corresponding volumes.

\begin{table}[!b!]	% Positionierung an der Unterkante einer Seite
	%\vspace{-0.2cm}	% Fließtext kann näher an die obere Kante der Tabelle herankommen, die Tabelle selbst verschieben sich nicht
		\centering
		\begin{threeparttable}
			\begin{tabular}{cc|c||cc|c}
				volume 1  &  volume 2  &  eff.~surface  &  volume 1  &  volume 2  &  eff.~surface  \\
				\hline
				skin  &  /  &  skin  &  /  &  skin  &  skin  \\
				skin$_\text{a}$  &  skin$_\text{b}$  &  skin$_\text{a}$\tnote{*}  &  skin$_\text{b}$  &  skin$_\text{a}$  &  skin$_\text{b}$\tnote{*}  \\
				border  &  /  &  border  &  /  &  border  &  /  \\
				border  &  skin  &  border  &  skin  &  border  &  skin  \\
				border$_\text{a}$  &  border$_\text{b}$  &  border$_\text{a}$  &  border$_\text{b}$  &  border$_\text{a}$  &  border$_\text{b}$  \\
				\hline
			\end{tabular}
			\begin{tablenotes}
				\footnotesize
				\item[*] If one of the volumes is the mother volume of the other volume, the effective surface is always the \texttt{G4LogicalBorderSurface} of the daughter volume, disregarding from which volume the photons are approaching the surface.
			\end{tablenotes}
		\end{threeparttable}
		%\vspace{-0.1cm}	% Tabelle rückt weiter nach unten (außer vor footnotes)
		\vspace{-0.2cm}	% Tabelle rückt weiter nach unten (außer vor footnotes)
	\end{table}
	%\vspace{-0.1cm}
	The following table summarises the effective \texttt{G4OpticalSurface}s, which are responsible for the surface properties, for all possible combinations of \texttt{G4OpticalSurface}s. Thereby, the photons cross the surface from volume 1 to volume 2, ``skin'' stands for a \texttt{G4LogicalSkinSurface} around the corresponding volume, ``border'' represents a \linebreak \texttt{G4LogicalBorderSurface} between both volumes (where the properties are specified for photons that approach the border from the volume corresponding to the column), and ``a'' or ``b'' distinguish different surfaces of the same type.

	\vspace{-0.1cm}
	The effect of these combinations has to be considered when creating \texttt{G4OpticalSurface}s. It is e.g.~a bad idea to use \texttt{G4LogicalBorderSurface}s to define the optical properties of the surface between outer cladding of an optical fibre and its surroundings. If the fibre is positioned in such a way that its cladding touches the outer surface of its mother volume (e.g.~for fibres directly at the surface of a scintillator tile or optical glue volume), a surface is formed between the fibre cladding and a volume different from the mother volume. This surface only possesses the desired properties, if a \texttt{G4LogicalSkinSurface} is used instead.
	\item
	\vspace{-0.1cm}
\begin{figure}[!b!]	% Positionierung an der Unterkante einer Seite
	\vspace{-0.3cm}	% Fließtext kann näher an die obere Kante der Bilder herankommen, die Bilder selbst verschieben sich nicht
	\centering
  \includegraphics[width=7.9cm]{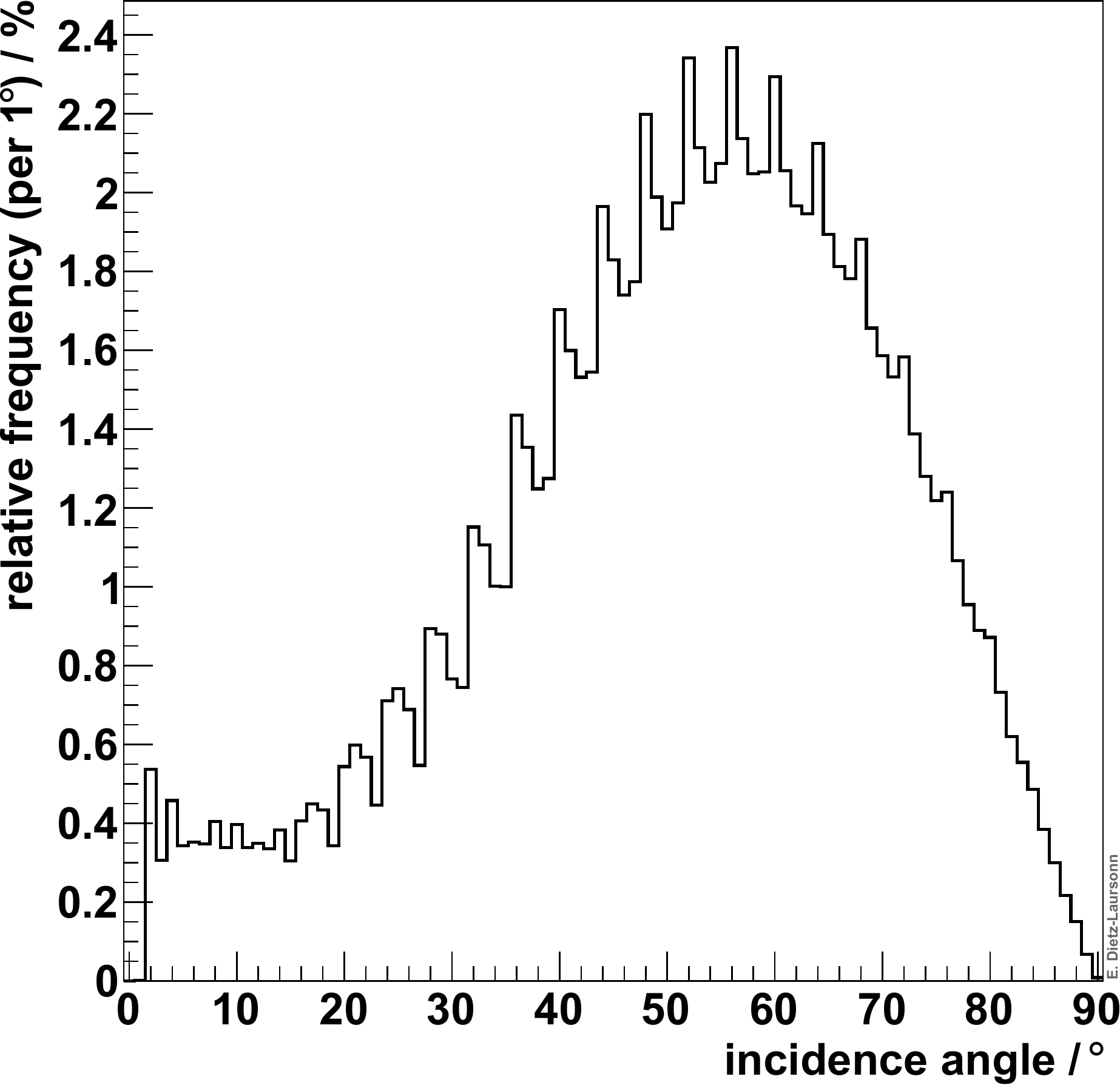}
	\hspace{0.5cm}
  \includegraphics[width=7.87cm]{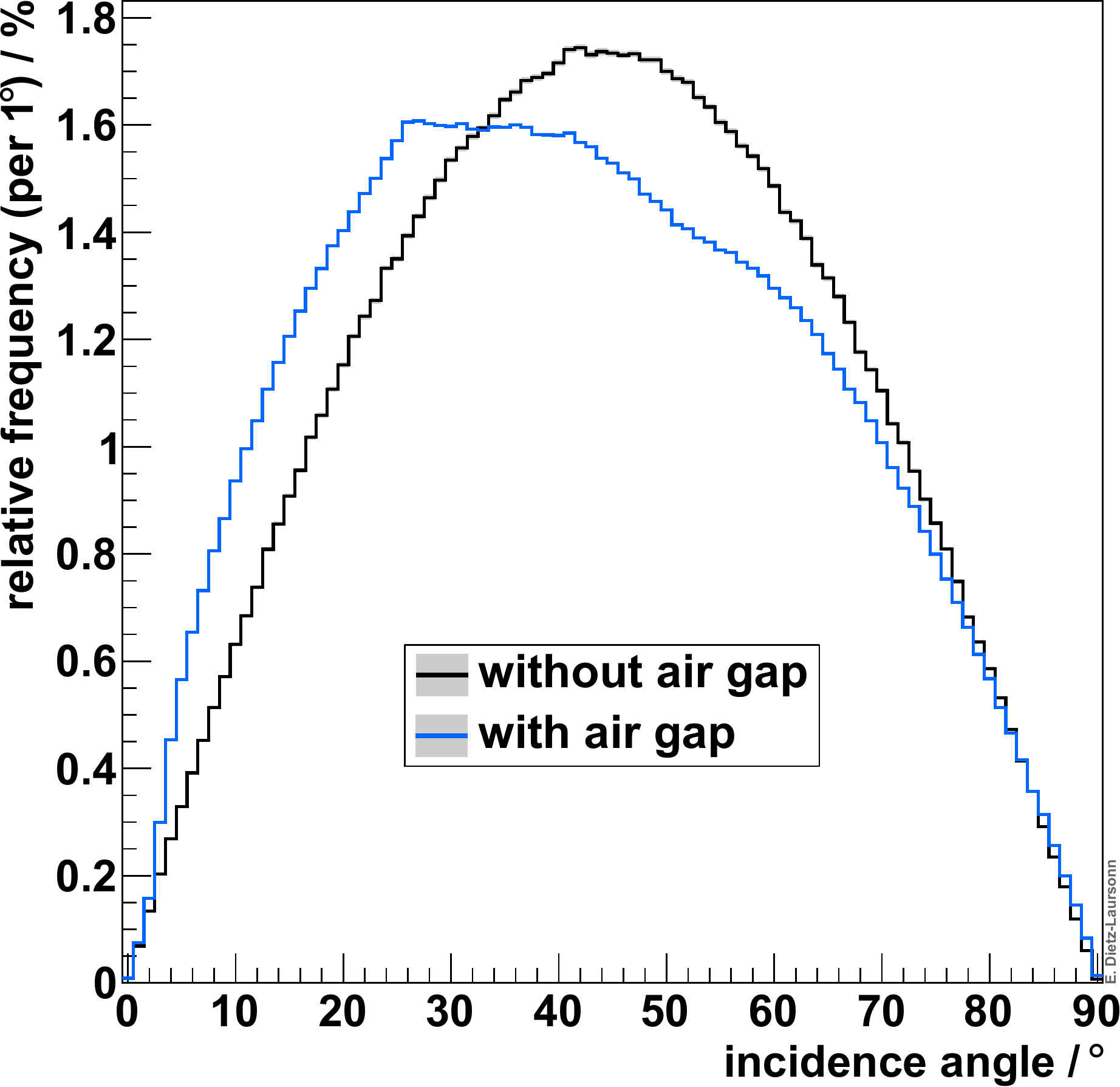}

	\vspace{-0.1cm}	% Abstand zwischen Bildern und Caption
	\caption{Simulated angular distribution of optical photons at the position of a photodetector for Teflon ``LUT-wrapping'' (left) and for a wrapping consisting of an additional volume with Teflon properties~(right). The distributions have been normalised to the number of entries.}
  \label{fig:validationOfMaterialPropertyInputAndPhysicsProcesses_LUT_angularDistribution}
	\vspace{-0.3cm}	% Bild rückt weiter nach unten (außer vor footnotes)
\end{figure}
	In Geant4, predefined properties from Look-Up Tables (LUT) can be used as alternative to manually defining the properties of a \texttt{G4OpticalSurface}. However, this method leads to unexpected results when using the surfaces to simulate wrappings. This is a known problem~\cite{Geant4HyperNews_LUT} and can e.g.~be shown using the example of a wrapped \unit[100$\times$100$\times$10]{mm$^3$} scintillator tile. The simulation was repeated three times: with a wrapping consisting of an additional volume with the material properties of Teflon (with and without a \unit[10]{$\upmu$m} air gap between scintillator and wrapping) and with a LUT named ``\texttt{groundteflonair}'' \texttt{G4LogicalBorderSurface} (representing the wrapping) between the scintillator and its mother volume. The scintillator was excited with simulated traversing muons and the incidence angle of the scintillation photons was determined on a \unit[3$\times$3]{mm$^2$} area at the centre of one of the lateral surfaces (the position of a photodetector). The results are presented in figure~\ref{fig:validationOfMaterialPropertyInputAndPhysicsProcesses_LUT_angularDistribution}.
It is obvious that even the underlying shape of the distribution completely differs for both wrapping types, i.e.~between the left and the right plot in figure~\ref{fig:validationOfMaterialPropertyInputAndPhysicsProcesses_LUT_angularDistribution}. The deviations between the two distributions in the right plot of figure~\ref{fig:validationOfMaterialPropertyInputAndPhysicsProcesses_LUT_angularDistribution} (\texttt{G4Wrapping} with and without air gap) are expected. They are a result of the different combinations of reflection types: Without air gap, there is only diffuse reflection at the Teflon surface. With air gap, there is also geometric reflection at the surface between scintillator and air.

Furthermore, for ``LUT-wrapping'' (left plot in figure~\ref{fig:validationOfMaterialPropertyInputAndPhysicsProcesses_LUT_angularDistribution}), the distribution features many sharp spikes. The reason for the deviation of the shape of the angular distribution is unknown. It is possible that the ``LUT-Teflon'' simply simulates different properties than the properties that have been used for the \texttt{G4Wrapping}. Unfortunately, it is not possible to verify this assumption, as it is not or not easily possible to determine the properties used by ``LUT-Teflon''.

Additionally, when inspecting the visualisation of single events of the simulation, it becomes clear that there is another problem. The ``LUT-wrapping'' is not fully opaque, but some photons escape from the edges of the scintillator volume. This is illustrated in figure~\ref{fig:validationOfMaterialPropertyInputAndPhysicsProcesses_LUT_escapingPhotons}.
\begin{figure}[!b!]	% Positionierung an der Unterkante einer Seite
	\vspace{-0.2cm}	% Fließtext kann näher an die obere Kante der Bilder herankommen, die Bilder selbst verschieben sich nicht
	\centering
  \includegraphics[width=7cm]{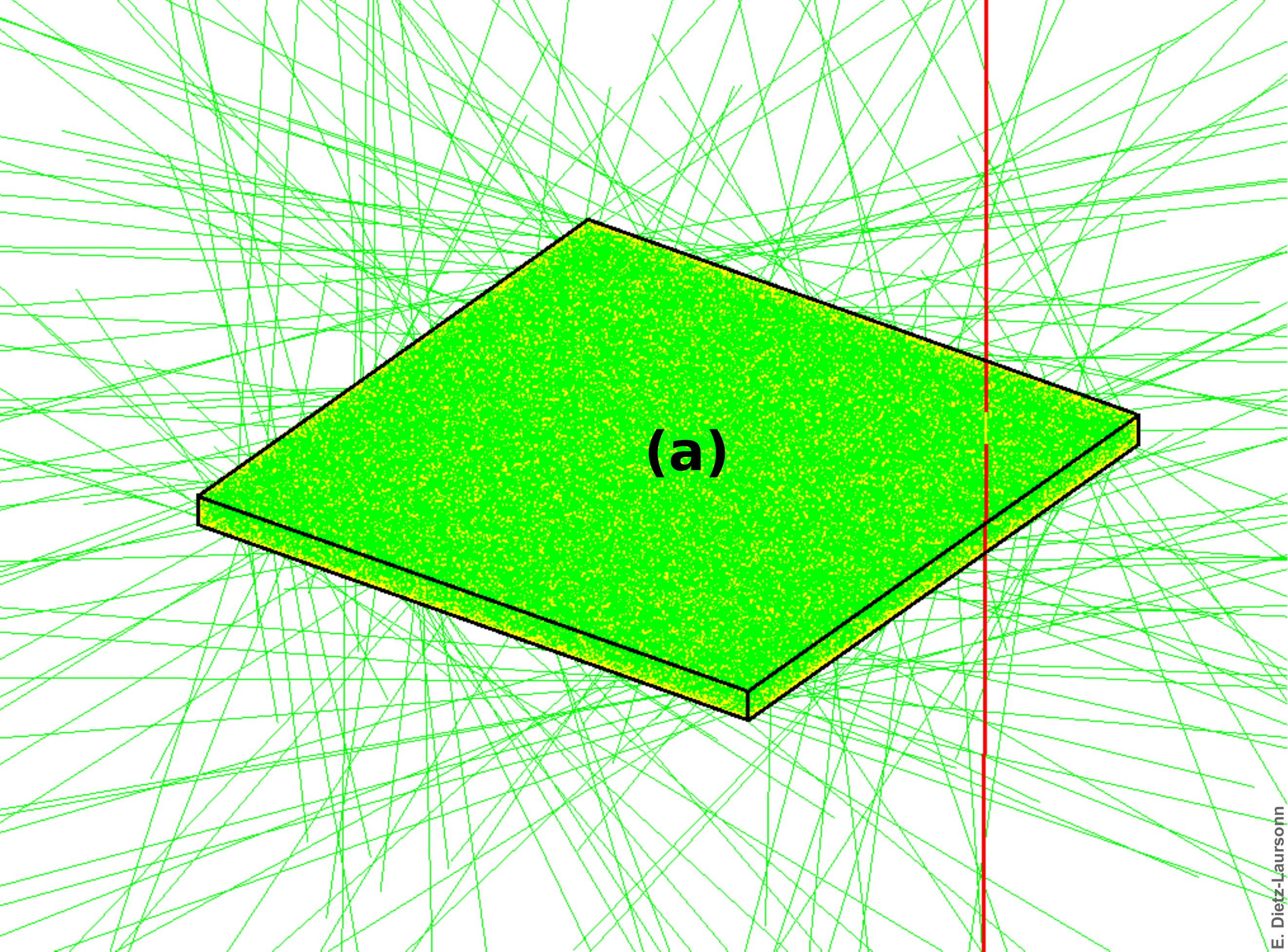}
	\hspace{1cm}
  \includegraphics[width=7cm]{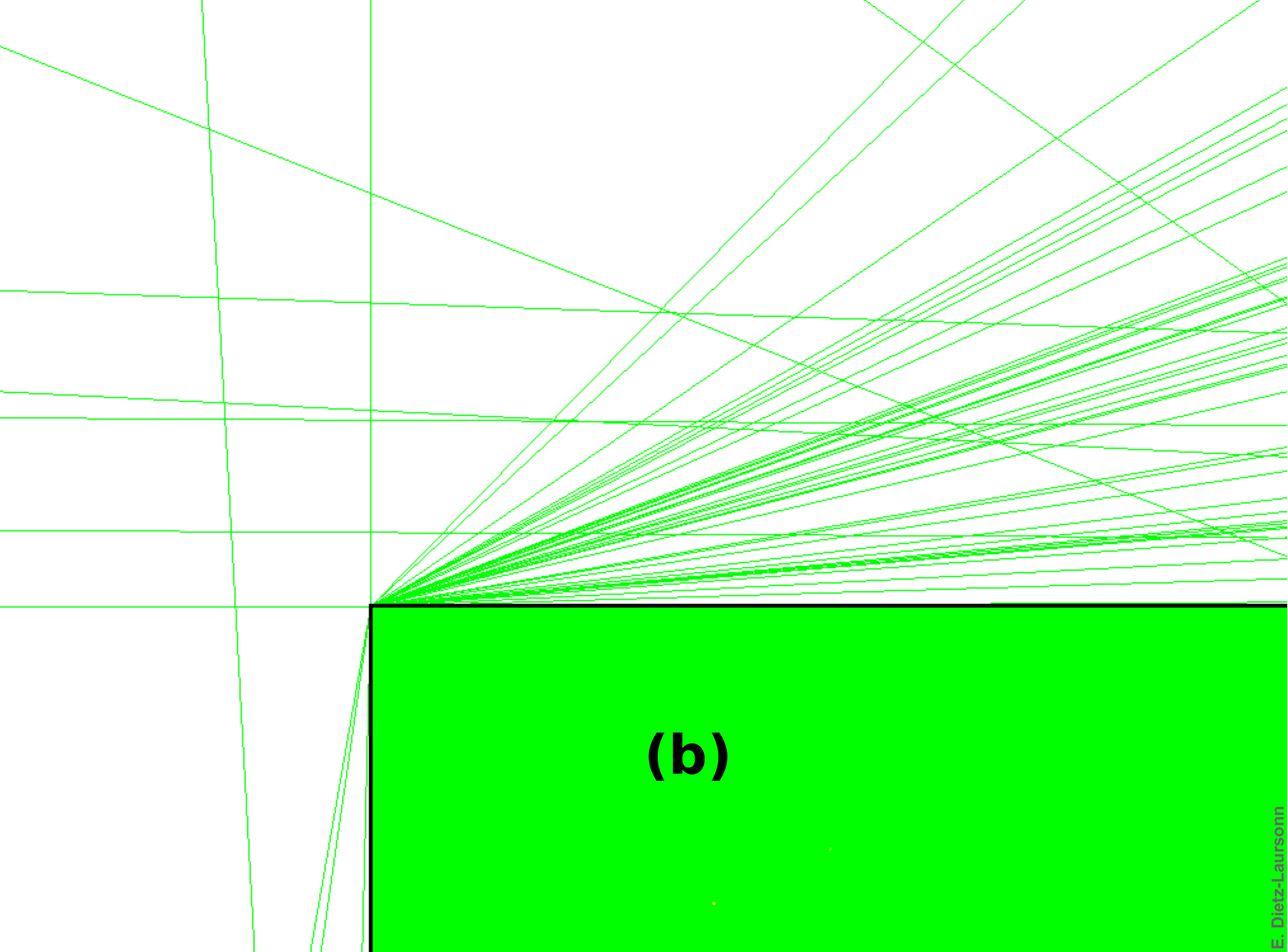}
	
	\vspace{-0.1cm}	% Abstand zwischen Bildern und Caption
	\caption{Escaping photons from a simulated scintillator tile with ``LUT-wrapping'': The scintillator tile is displayed in a slanted view from above (a) and zoomed into a side view (b). The muon is displayed in red and the photons are green. The scintillator tile seems to be homogeneously filled with green because of the large number of photons and reflections. If the wrapping was opaque, no green lines should be outside the scintillator volume. Obviously, some photons leave this volume, preferentially at the edges. In total, ca.~\unit[2.6]{\%} photons escaped in this event.}
  \label{fig:validationOfMaterialPropertyInputAndPhysicsProcesses_LUT_escapingPhotons}
	\vspace{-0.2cm}	% Bild rückt weiter nach unten (außer vor footnotes)
\end{figure}

Apart from the deviations in the angular distribution, the spikes in the angular distribution of the ``LUT-wrapping'' as well as the ``escaping'' photons are definitely unexpected and non-physical. Altogether, using LUT surfaces is therefore not recommended, at least for the simulation of wrappings.
	\item
	In case of ``dielectric-metal'' \texttt{G4OpticalSurface}s (e.g.~for the simulation of a mirror or a mirrored fibre end face), a modified Fresnel equation is used to calculate the reflectivity of the surface. This is a known issue in Geant4~\cite{Geant4HyperNews_Fresnel}. This modified equation only considers the complex refractive index of the ``dielectric-metal'' \texttt{G4OpticalSurface}, rather than the refractive indices of both materials that form the surface. If neither a complex refractive index nor the reflectivity is specified for ``dielectric-metal'' \texttt{G4OpticalSurface}s, the reflectivity is \unit[100]{\%} by default. Besides the fact that only considering one refractive index is physically not absolutely correct, this does not make any sense for surfaces between two dielectrics. The complex refractive index should therefore only be used for \texttt{G4OpticalSurface}s between a dielectric and a metal, where in the best case the refractive index of the dielectric is close to 1. Another possibility is to pass an effective complex refractive index (considering the refractive indices of both materials forming the surface) to the \texttt{G4OpticalSurface}.
\end{itemize}

%%%%%%%%%%%%%%%%%%%%%%%%%%%%%%%%%%%%%%%%%%

\subsection{Optical Property Variables}

The optical property variables are used in Geant4 to specify the properties of the optical physics processes during the simulation. They can have different meanings, depending on the object they are referred to:
\begin{itemize}
	\item
	The refractive index is not a surface property but a material property. Therefore, assigning it to a \texttt{G4OpticalSurface} has no effect. However, exceptions are \texttt{G4OpticalSurface}s with the surface finishes named ``-\texttt{backpainted}''. Here, a refractive index has to be defined, which is interpreted as the refractive index of the medium in the gap between the painted volume and the paint.
	\item
	In contrast, the complex refractive index is not treated as a material property but as a surface property. Therefore, it only has an effect when being specified for \texttt{G4OpticalSurface}s. In particular, assigning a complex refractive index to a material does not result in an implicit specification of the attenuation length spectrum. This treatment of the complex refractive index is physically reasonable, as all optical processes besides reflection probability and absorption (which is directly specified for materials) only depend on the real part of the refractive index (c.f. e.g.~\cite{The_Feynman_Lectures_on_Physics__31}).
	\item
	For direct specification of the reflectivity of \texttt{G4OpticalSurface}s, two optical property variables can be set and have to be distinguished. One is the \texttt{REFLEC\-TIVITY} variable. In case of a surface between a dielectric and a metal, it does exactly what its name suggests: The optical photons are reflected with the probability that has been set via the \texttt{REFLECTIVITY} variable (i.e.~bypassing the Fresnel equations). Not reflected photons are absorbed. Strictly speaking, this is a discrepancy with respect to the behaviour of ``realistic'' surfaces, which do not absorb photons. Photons that are not reflected at such a surface enter the next volume and undergo absorption corresponding to the attenuation length of the material, which is typically very short for optical photons in metals.
		
	However, in case of a surface between two dielectrics, the result is not as the variable name suggests: The photons are absorbed with a probability of $\unit[100]{\%} - r$, where $r$ is the specified ``reflectivity'' value. The remaining photons are then reflected/refracted according to the Fresnel equation. Thus, in this case, the \texttt{REFLECTIVITY} variable simulates something like absorbing dirt on the surface. To specify the reflectivity of a surface between two dielectrics, the \texttt{TRANSMITTANCE} variable has to be used, where the reflectivity $R$ derives from the transmittance $T$ by $R = \unit[100]{\%} - T$. One should also note that Geant4 does not perform any optical process at a surface between two dielectrics that consist of the same material.
	\item
	For WLS materials, two different attenuation length variables can be specified: In addition to the \texttt{ABSLENGTH} variable, which defines the usual absorption of optical photons, the \texttt{WLSABSLENGTH} variable corresponds to absorption which triggers the WLS process. Both variables can be combined for one material, e.g.~to simulate WLS material with a limited WLS absorption range in combination with an absorption spectrum for the created WLS photons. If both spectra overlap, both are taken into account, i.e.~the process with the shorter attenuation length is more likely in the overlap region.
\end{itemize}

\section{Summary}

There are several peculiarities regarding the simulation of the optical physics with Geant4. They are caused by the fact that the optical physics process in Geant4 is designed to be very flexible and extensive.
Dealing with them in the wrong way can have a huge influence on the results of the simulation and even lead to non-physical results.
These peculiarities, which the user has to be aware of in order to avoid mistakes, have been summarised in this paper.
This was done to assist the user, as these peculiarities are not always easy to recognise and thus can be a serious problem.

	% Danksagung
\section*{Acknowledgements}

We gratefully acknowledge financial support from the "Helm\-holtz-Ge\-mein\-schaft Deut\-scher For\-schungs\-zen\-tren" (HGF)
under contract number HA-101 (Physics at the Terascale) and from the "Bun\-des\-mi\-ni\-ste\-ri\-um f\"ur Bil\-dung und
For\-schung" (BMBF) under contract number 05H12PA1.
Special thanks go to the people of the Geant4 Hypernews, particularly Peter Gumplinger, for many helpful comments and fruitful
discussion.

	% Anhang: Bibliographie
	\printbibliography

\end{document}